\documentclass[shortnote, preprint]{jpsj9}
\usepackage{txfonts}
\usepackage{graphicx} 
\usepackage{epstopdf}

\title{Hydrostatic pressure studies on non-superconducting UTe$_2$}

\author{M. O. Ajeesh\thanks{ajeesh@lanl.gov}, Joe D. Thompson, Eric D. Bauer, Filip Ronning, Sean~M.~Thomas, and Priscila F. S. Rosa}
\inst{Los Alamos National Laboratory, Los Alamos, NM 87545, USA}

\abst{We report the temperature--pressure phase diagram of a UTe$_2$ single crystal that does not undergo a bulk superconducting transition but shows filamentary superconductivity with a critical transition temperature of 1~K at ambient pressure. Electrical-resistivity measurements reveal that the evolution of the filamentary superconducting state under pressure resembles the behavior observed in previous reports on bulk superconducting samples. AC calorimetry, however, does not show evidence for either bulk superconductivity or magnetism for pressures up to 1.6~GPa. Our results highlight the role of inhomogeneity in chemical-vapor-transport-grown UTe$_2$ samples and serve as a cautionary tale when probing electrical resistivity alone.
}

\begin{document}
\maketitle

Following the recent discovery of odd-parity superconductivity in UTe$_2$~\cite{Ran19}, this puzzling material has been heavily investigated as a chiral superconductor candidate (see reviews~\cite{Aoki22,Lewin23}). UTe$_2$ also presents many intriguing properties including multiple pressure-induced superconducting (SC) phases that give way to antiferromagnetism above 1.4~GPa~\cite{Braithwaite19, Thomas20,Li21,Knafo23}. A distinct field-reinforced SC phase emerges when magnetic fields are applied along the $b$ axis. Further, a reentrant SC phase appears within the field-polarized phase for fields along a specific angular range in the $b-c$ plane~\cite{Ran19b,Lewin23}.

The precise nature of the superconducting order parameter in UTe$_2$ remains an open question even at zero pressure and field. While recent experimental studies suggest that the ambient-pressure SC phase does not intrinsically break time-reversal symmetry and is likely non-chiral~\cite{Ajeesh23,Azari23,Theuss23}, the symmetry of its order parameter is yet to be unambiguously determined. Far less is understood about the pressure- and magnetic field-induced superconducting and magnetic phases. 

A major hurdle in understanding the intrinsic properties of UTe$_2$ has been the sample dependence of the experimental results. The SC transition temperature, $T_c$, the residual specific heat coefficient in the SC state, and the residual resistance ratio (RRR) drastically vary among crystals grown in different conditions despite having similar structural properties~\cite{Rosa22,Haga22,Weiland22}. Samples grown via the chemical vapor transport (CVT) technique exhibit systematic changes in $T_c$ and in RRR as a function of the growth temperatures, and superconductivity abruptly disappears for samples grown beyond certain upper and lower growth-temperature limits~\cite{Rosa22,Weiland22}. Notably, samples grown via the molten-salt flux (MSF) technique are of higher quality, as evidenced by higher $T_c$ and RRR values and the presence of quantum oscillations~\cite{Sakai22}. 

Recent high-field studies on MSF-grown UTe$_2$ samples report that the field-reinforced SC phase is sensitive to disorder whereas the reentrant SC phase is much more robust~\cite{Wu23}. Remarkably, high-field studies on CVT-grown UTe$_2$ reported that the reentrant SC phase occurs even in samples that are not superconducting at zero field, albeit in a narrower angular range compared to zero-field superconducting UTe$_2$~\cite{Frank23}. These observations point to the different susceptibility of the various high-field SC phases to disorder. Given the significant sample dependence of the ambient-pressure superconducting phase, investigating the effect of disorder on the pressure- and field-induced phases could provide insights into their microscopic origin.
\begin{figure}[t]
\centering
\includegraphics[width=\linewidth]{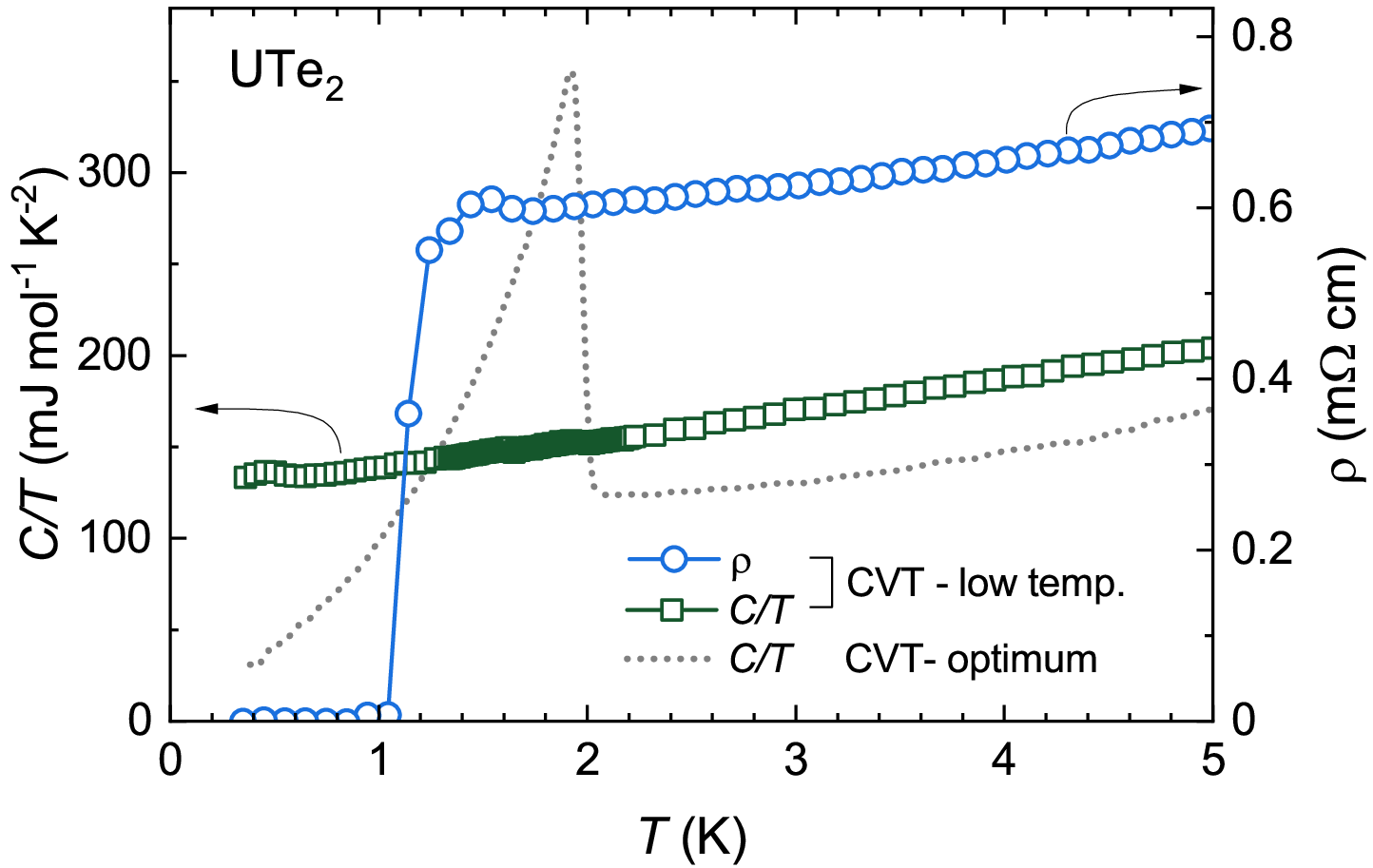}
\caption{Temperature dependence of the heat capacity (left axis) and electrical resistivity (right axis) of a UTe$_2$ sample grown at lower temperature via chemical vapour transport (CVT) method. Dotted line represents the $C/T$ vs. $T$ for a UTe$_2$ sample, grown via optimum CVT conditions, showing bulk superconductivity.}
\vspace{-1cm}
\label{Fig1}
\end{figure}

Here we address the open question of whether the pressure-induced superconducting and magnetic phases in UTe$_2$ survive in samples that are not superconducting at ambient pressure (hereon called NSC).
Single crystals of UTe$_2$ were grown using the CVT method~\cite{Rosa22}. Electrical resistivity (standard four-probe method) and heat capacity (thermal-relaxation technique) at ambient pressure were measured using a Physical Property Measurement System (Quantum Design). Electrical-resistivity and ac-calorimetry~\cite{Sullivan68} measurements under hydrostatic pressure were performed using a piston-clamp pressure cell with Daphne 7373 oil as the pressure medium. Pressure was determined using a Pb manometer. Pressure-dependent measurements at low temperatures were carried out in an adiabatic demagnetization refrigerator. 
\begin{figure*}[t]
\centering
\includegraphics[width=1\linewidth]{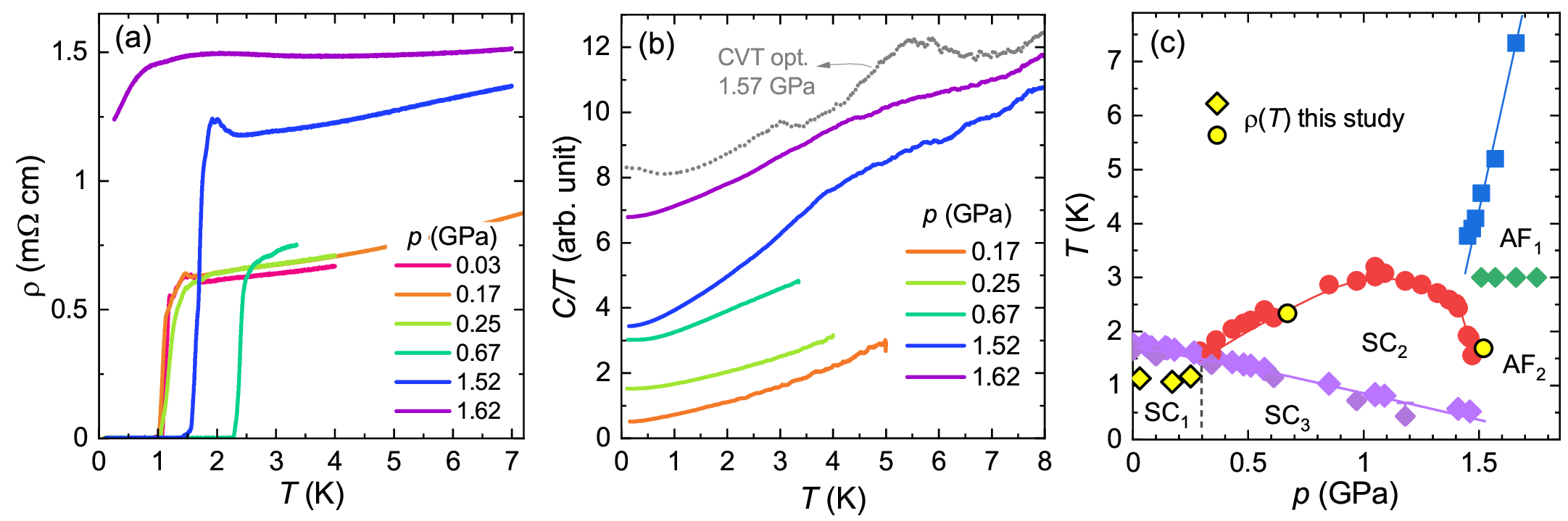}
\caption{Temperature dependence of (a) electrical resistivity and (b) heat capacity measured on the same NSC UTe$_2$ sample for several applied pressures. The dotted curve in (b) is the heat capacity of an optimum-CVT grown UTe$_2$ sample measured at 1.57~GPa. (c) Temperature--pressure phase diagram adapted from Ref.~\cite{Thomas20,Thomas21}. The yellow symbols mark the superconducting transition temperatures obtained from the resistivity data in the present study.}
\vspace{-0.5cm}
\label{Fig2}
\end{figure*}

To investigate the pressure evolution of the properties of a NSC UTe$_2$ sample, we chose a crystal grown via CVT in the cold end of a quartz tube maintained at a temperature gradient between 775$^0$C and 685$^0$C. Under these growth conditions, UTe$_2$ crystals have significant uranium vacancies (up to 5~\%), a reduced unit cell volume, and larger atomic displacement parameters compared to bulk SC samples \cite{Weiland22}.  Figure~\ref{Fig1} shows the temperature dependence of the electrical resistivity (right axis) and heat capacity (left axis) of the investigated sample at ambient pressure. Electrical resistivity ($\rho(T)$) shows a zero-resistance transition around 1 K; however, there is no feature in heat capacity ($C/T$) in the corresponding temperature scale. For a comparison, the dotted line in Fig.~\ref{Fig1} shows a large heat capacity anomaly at the SC transition of a sample grown via CVT in optimized conditions. This results indicates that a very small fraction of the investigated NSC sample undergoes a SC transition.

The temperature dependence of electrical resistivity and heat capacity of the NSC UTe$_2$ sample under several applied pressures is presented in Fig.~\ref{Fig2}. The SC transition observed in $\rho(T)$ first shifts to lower temperatures and then to higher temperatures before decreasing again with increasing pressure, similar to previously reported pressure dependence for bulk-SC UTe$_2$ samples. However, temperature-dependent heat capacity data, $C/T(T)$, do not show any anomaly corresponding to the SC transitions seen in $\rho(T)$. This result clearly shows that the pressure-induced SC transition is also limited to a small volume fraction, and there is no bulk SC under pressure. In Fig.~\ref{Fig2}c, the pressure dependence of $T_c$ determined from $\rho(T)$ is mapped onto the temperature--pressure phase diagram of bulk-SC samples reported previously~\cite{Thomas20, Thomas21}. Here, $T_c$ is the midpoint of the drop in $\rho(T)$ at the SC transition. Notably, $T_c$ of the pressure-induced SC phase above $0.3$~GPa is comparable to that for bulk-SC samples despite having a much lower $T_c$ below $0.3$~GPa. Further, there is no SC transition in $\rho(T)$ for $p=1.62$~ GPa, suggesting a similar critical pressure as in the bulk SC samples. 

Above 1.4 GPa, bulk SC samples give rise to two magnetic transitions evidenced by electrical-resistivity and heat-capacity anomalies~\cite{Thomas20} (the dotted curve in Fig.~\ref{Fig2}(b) shows the heat capacity of an optimum-CVT grown UTe$_2$ sample measured at 1.57 GPa, evidencing two magnetic transitions). However, in the NSC sample investigated here, evidence for a magnetic transition is less clear. A weak, broad hump in $C/T(T)$ is observed at the highest pressures in the temperature range $3-6$~K (Fig.~\ref{Fig2}b), which is possibly due to short-range magnetic correlations. In such a scenario, uranium vacancies would be responsible for the failure to develop long-range magnetic order under pressure. 

In spite of the significant susceptibility to disorder of both SC1 and magnetic phases, our data reveal a comparatively increased robustness of $T_c$ in the pressure-induced SC2 phase, a finding with numerous ramifications. First, the qualitative difference in the response of SC1 and SC2 to disorder corroborates that the two SC phases host distinct SC gap functions. Second, the robustness of SC2 is at odds with a recent nuclear magnetic resonance (NMR) report that argues for a close connection between SC2 and the field-reinforced SC state, which is sensitive to disorder~\cite{Kinjo23,Wu23}. Third, the robustness of SC2 could in principle be explained by the presence of a fully gapped and isotropic gap function; however, this conclusion is inconsistent with the absence of a Knight shift drop at $T_c$ under pressure~\cite{Kinjo23}. We therefore speculate that the increased robustness of SC2 comes from the generalization of Anderson's theorem wherein an unconventional SC state with multiple internal degrees of freedom (e.g., orbitals or sublattices) is not affected by nonmagnetic disorder~\cite{Andersen20}. Further experimental and theoretical investigations of the SC order parameter of UTe$_2$ under pressure are required to test this scenario.    

\begin{acknowledgments}
This work was supported by the U.S. Department of Energy, Office of Basic Energy Sciences “Quantum Fluctuations in Narrow Band Systems” program. M.O.A. acknowledges funding from the Laboratory Directed Research \& Development Program.  
\end{acknowledgments}

\end{document}